\documentstyle[natbib,satellites,mycommands,times,graphicx]{aipproc}

\newcounter{Zaehler}
\newenvironment{Aufzaehlung}{\setcounter{Zaehler}{1}%
\begin{list}{\arabic{Zaehler}.}{%
\usecounter{Zaehler}\labelwidth1em\labelsep0.5em
\leftmargin\labelwidth\advance\leftmargin by \labelsep
\parsep0pt\itemsep0.1\baselineskip\topsep0pt\partopsep0pt%
}}{\parskip0pt\end{list}}

\newcommand{\altaffilmark}[1]{\mbox{${}^{#1}$}}

\renewcommand{\ca}{\mbox{$\sim$}}
\renewcommand{\mal}{\mbox{$\times$}}

\newlength{\figurewidth}
\begin{document}

\title{Disappearing Pulses in Vela X-1}

\author{%
  P. Kretschmar\altaffilmark{1,2}, 
  I. Kreykenbohm\altaffilmark{2}, 
  J. Wilms\altaffilmark{2}, 
  R. Staubert\altaffilmark{2}, \\
  W. A. Heindl\altaffilmark{3}, 
  D. E. Gruber\altaffilmark{3}, 
  R. E. Rothschild\altaffilmark{3}%
}
\address{\altaffilmark{1}\,INTEGRAL Science Data Centre,
  Ch. d'Ecogia 16, 1290 Versoix, Switzerland\\
\altaffilmark{2}\,Institut f\"ur Astronomie und Astrophysik --
  Astronomie, Waldh\"auser Str. 64, D-72076 T\"ubingen, Germany\\
\altaffilmark{3}\,CASS, University of California at San Diego, La
  Jolla, CA 92093, U.S.A.}

\maketitle

\begin{abstract}
  We present results from a 20\,h RXTE observation of \mbox{Vela X-1},
  including a peculiar low state of a few hours duration, during which
  the pulsation of the X-ray emission ceased, while significant non-pulsed
  emission remained. This ``quiescent state'' was preceded by a ``normal 
  state'' without any unusual signs and followed by a ``high state'' of
  several hours of increased activity with strong, flaring pulsations.
  While there is clear spectral evolution from the normal
  state to the low state, the spectra of the following high state are
  surprisingly similar to those of the low state.
\end{abstract}

\section*{Introduction}
\vela (4U\,0900--40) is an eclipsing high mass X-ray binary
consisting of the 23\,\mbox{$M_{\odot}$} B0.5Ib supergiant \mbox{HD\,77581}
and a neutron star with an orbital period of 8.964\,d
and a spin period of about 283\,s 
\citep[and references therein]{vanKerkwijk:95}.
The persistent X-ray flux from the neutron star is known to be very 
variable exhibiting strong flares and low states. \cite{Inoue:84} and
\cite{Kreykenbohm:99} have observed low states of near quiescence where
no pulsations were seen for a short amount of time. Before or after 
these low states normal pulsations were observed.
%
During an observation of \vela for 12 consecutive orbits in January 1998 by
the Rossi X-ray Timing Explorer (\xte), we have by chance observed such a
quiescent state for the first time from the beginning to the end, preceded
and followed by the usual pulsations.



\section*{Lightcurves and pulse profiles}

\begin{figure}
  \centerline{\includegraphics[width=1.0\figurewidth]{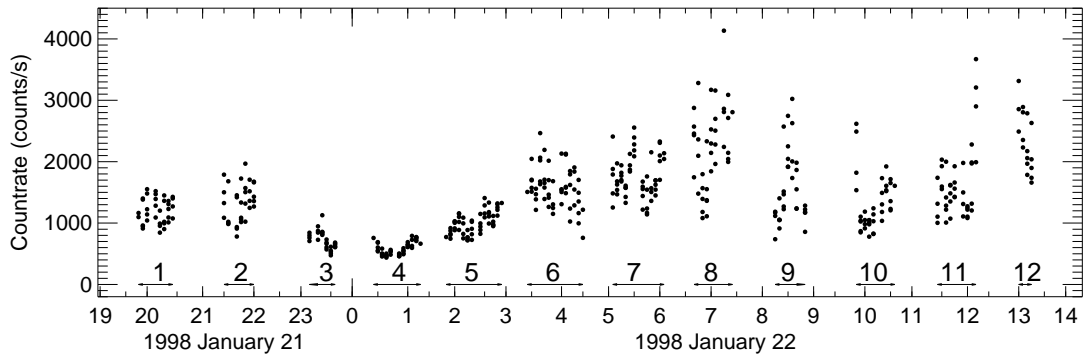}}
  \vspace*{4mm}
  \caption{\label{fig:TotalLC}Light curve of the complete observation. 
    The individual \xte orbits are indicated below the light curve.}
\end{figure}
\begin{figure}
  \centerline{\includegraphics[width=1.0\figurewidth]{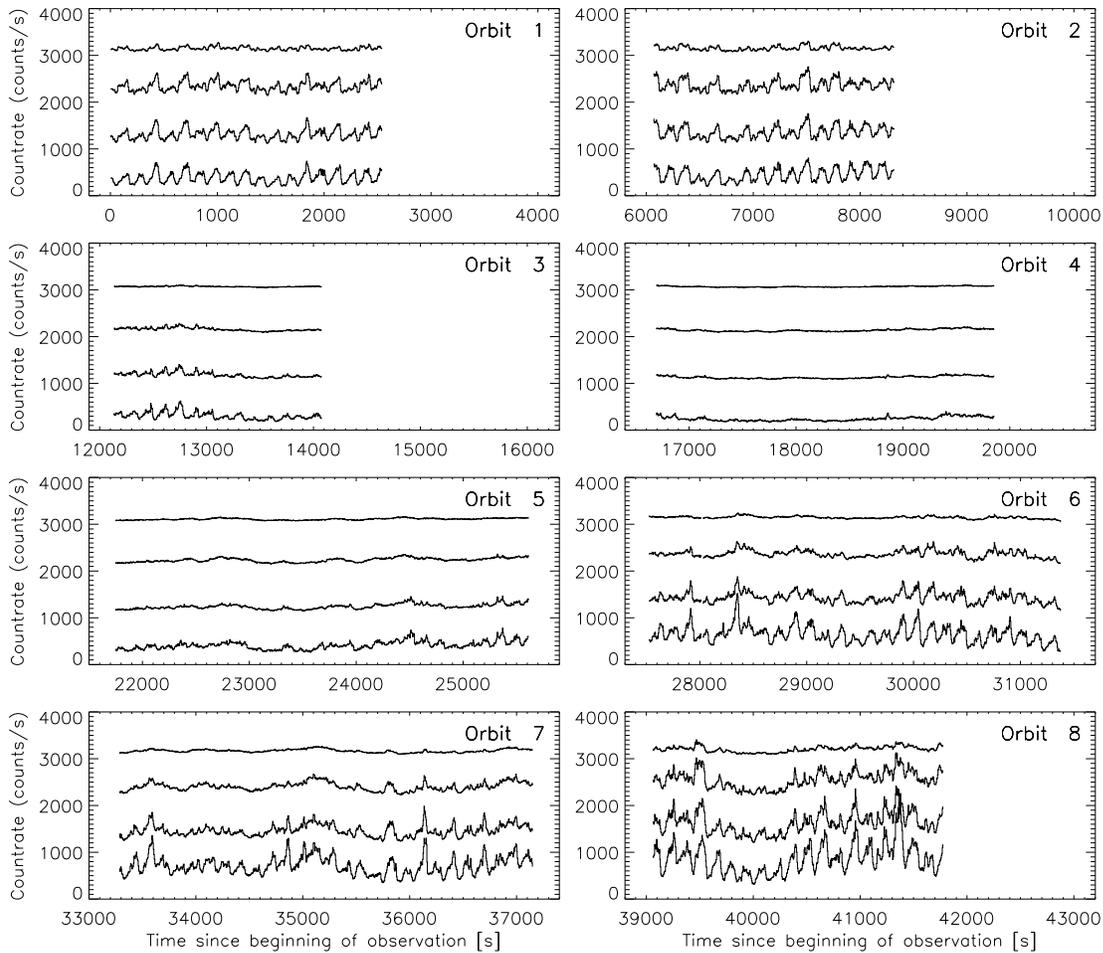}}
  \vspace*{2mm}
  \caption{\label{fig:ColorLC}Lightcurves in four energy bands 
    for the individual orbits~1 to~8. The energy bands are the same as for
    the pulse profiles (see \Fig{fig:PP}).  The lightcurve for the highest
    energy band is plotted at the bottom with its actual count rate, the 
    others have been shifted upwards by 1000, 2000 and 3000 counts/s
    respectively.}
\end{figure}

\begin{figure}
  \centerline{\includegraphics[width=\figurewidth]{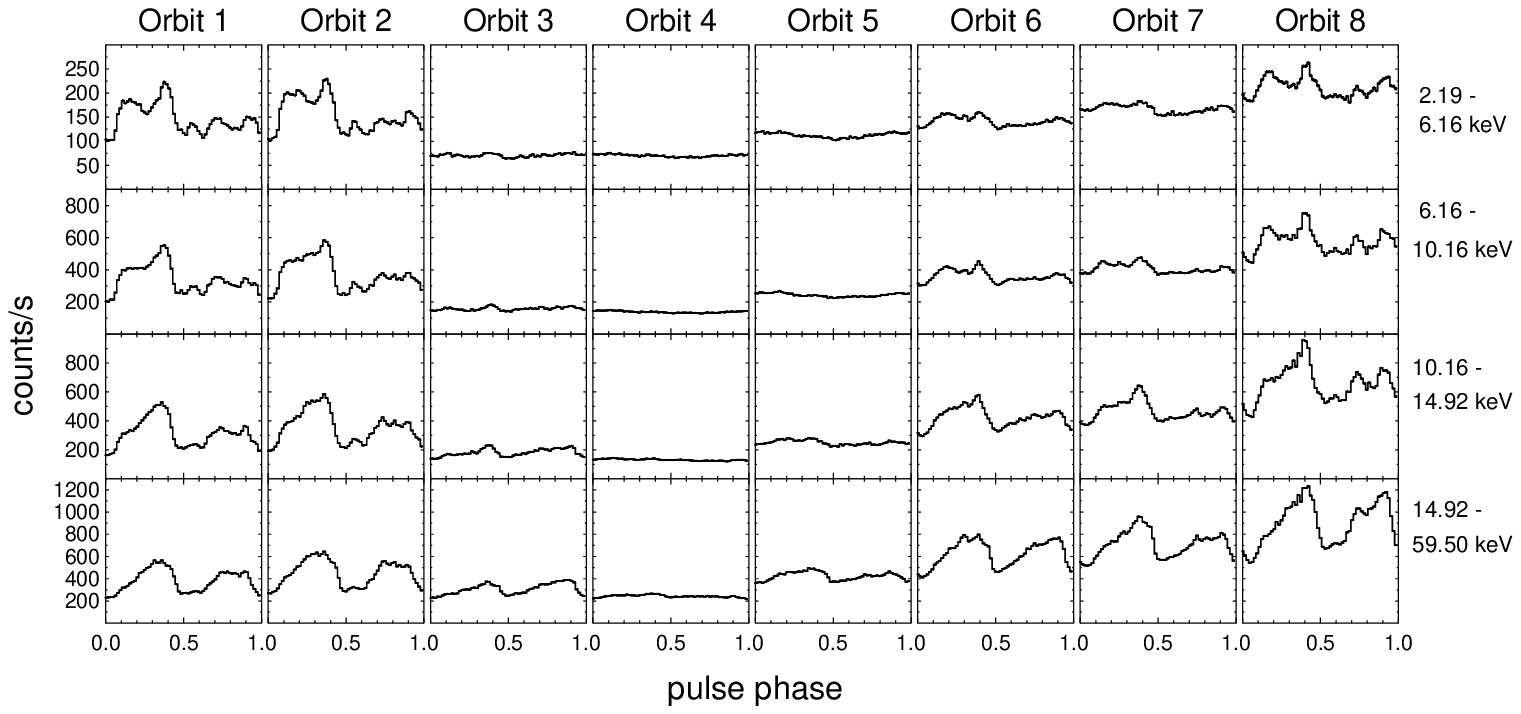}}
  \vspace*{2mm}
  \caption{Evolution of the pulse profile for the individual orbits~1 to ~8. 
    The energy bands are the same used in \Fig{fig:ColorLC}.  The
    lightcurves were folded with a fixed period (283.3\,s) and a common zero
    time.}
  \label{fig:PP}
\end{figure}

As \Fig{fig:TotalLC} demonstrates, the source flux suddenly decreased
between orbits~2 and~3, reaching its minimum during orbit~4. At the same
time \emph{the source pulsations decreased strongly, while significant
  non-pulsed source flux remained}. This is shown in detail in
\Fig{fig:ColorLC}.  The pulsed fraction 
decreased from 
30\%--50\%, depending on the energy band, to 7\%--9\%.
Note that even at the lowest state, the overall source
flux was $>$5 times the predicted background level in the energy range
used.

\Fig{fig:PP} presents the pulse profiles obtained from the individual
orbits~1 to~8.  The profiles of the first two orbits correspond to the
well-known, complex shape usually obtained when integrating over many
pulse periods, with a clear transition from a five-peaked profile at
low energies to a double-peak structure at high energies 
\citep{Raubenheimer:90}.
In contrast, orbits~3 to~5 show much less pronounced profiles, with
the profile of orbit~4 being essentially flat. The pulse profile of
the source during ``recovery'' (orbits~6 \& 7) is similar to those
observed before the low state at higher energies but much less
pronounced at energies $<$10\,keV.

\section*{Hardness ratios and Spectra}

\begin{figure}
  \centerline{\includegraphics[width=1.05\figurewidth]{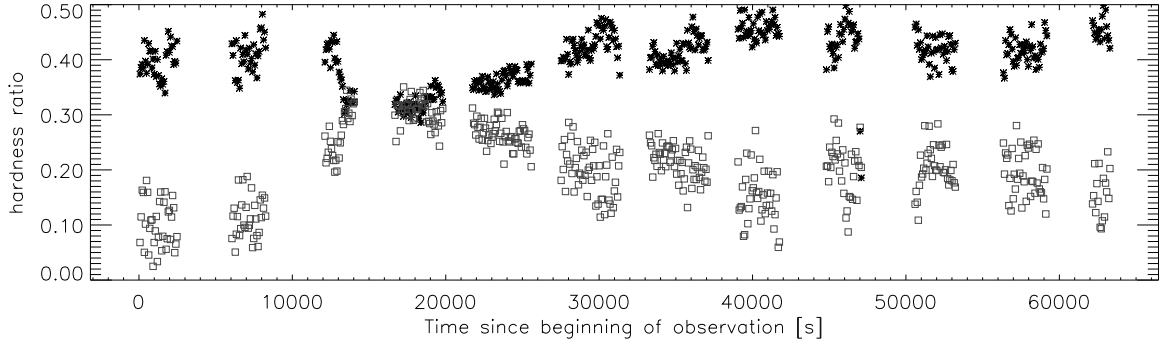}}
  \caption{Evolution of the spectral hardness over the complete observation.
    The upper curve (stars) displays the ratio of energy band 2 to band 1,
    the lower curve (open squares) the ratio of band 4 to band 3.}
  \label{fig:Hardness}
\end{figure}

\begin{figure}
  \centerline{\includegraphics[width=1.00\figurewidth]{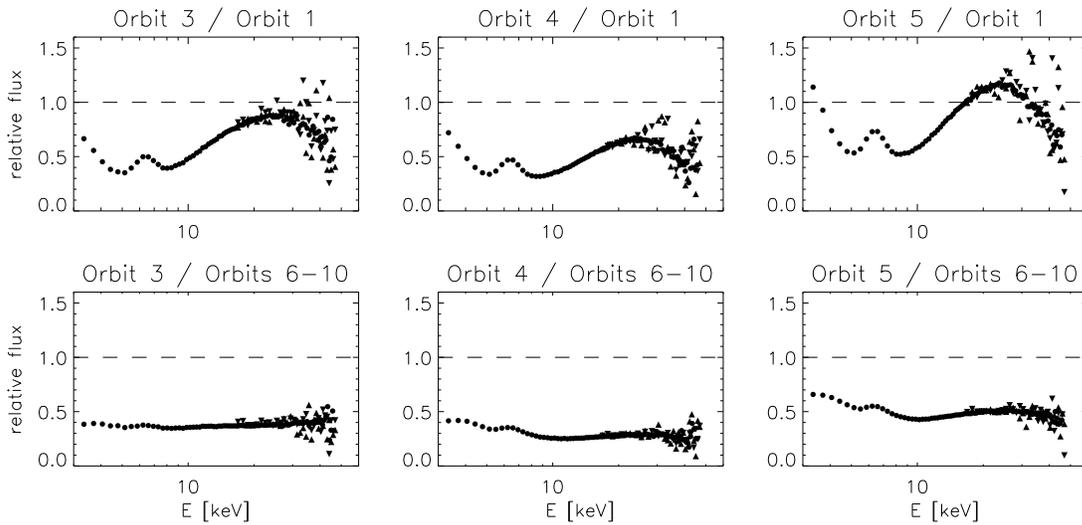}}
  \caption{Quotient spectra of the three ``low-state'' orbits, with
           \pca data plotted as filled circles and \hexte data as triangles: 
           1. upper panels: ratio to the spectrum of the normal state before 
              the low state, 
              2. lower panels: ratio to the average spectrum of the
              active high state following the low state.}
  \label{fig:Quotspec}
\end{figure}

\Fig{fig:Hardness} shows the evolution of the spectral hardness, both at
energies up to 10\,keV (energy band 2 vs.\ band 1) and at energies beyond
10\,keV (energy band 4 vs.\ band 3).  The hardness ratios were calculated
using $(H-S)/(H+S)$ where $H$ and $S$ are the fluxes in the hard and
soft band respectively. There are three apparent properties of this plot:
\begin{Aufzaehlung}
\item \emph{The two hardness ratios are very clearly anticorrelated.} 
\item \emph{The disappearence of the pulses in orbit~3 goes hand in hand
        with an abrupt spectral change.}
\item The reemergence of pulsations is accompanied by a 
  ``normalization'' of the hardness ratios, \emph{but during the high state 
  the spectrum stays significantly harder than in the normal state at the 
  beginning.}
\end{Aufzaehlung}
The strong spectral changes at the onset of the low state are also apparent
in the quotient spectra shown in \Fig{fig:Quotspec}. There are clear
indications for strong absorption and at the same time increased flux in the
iron line and a soft excess at the lowest energy range. 

\emph{In contrast
  there is little change in the global spectral shape as the source begins
  to pulsate again.  Except for a slight soft excess, the spectrum during
  the low state is rather well described by simply scaling the spectrum of
  the following flaring state.}

Attempts to fit the spectra turned out to be quite difficult, even allowing
for the known complexity of the \vela continuum. Detailed results have been
presented on two posters at the \mbox{X-ray'999} conference in Bologna, the
following paragraphs summarize the results.

We used a partial covering model with two additive components, using
the same continuum, one heavily absorbed and one scattered unabsorbed into
the direction of the observer. The continuum 
used was the NPEX
(Negative Positive EXponential \citealt{Mihara:Dr}). This model had to be
further modified by an additive iron line and two coupled cyclotron lines.
For the high state a cyclotron line at \ca 55\,keV is required by the data,
a second feature at \ca 21\,keV may be due to uncertainties in the \pca
response matrix. 

Modeling the spectra of the individual orbits in the first part of our
observation, we found the N\i{H} value of the absorbed component
varying between 40\mal 10\e{22}\,cm\e{-2} and 230\mal
10\e{22}\,cm\e{-2} with a clear maximum during orbit~4. The relative
importance of the scattered component appears also to be maximal
during the low state. There is no clear correlation of the other
continuum parameters with source flux, but their values are quite
different before and after the low state.

\section*{Discussion}
The results of spectral fitting are somewhat in contrast with the finding
above that the global spectral shape remains more or less constant after
orbit~3. Within the framework of our spectral model this similarity is
obtained by parallel changes in the column depth and in the spectral
continuum. Further analysis will have to show if this is an artefact or
reality.


A possible scenario to explain the disappearing pulses is that a very thick
blob in the surrounding stellar wind -- which is known to be clumpy
\cite[e.g.,][]{Nagase:86} -- temporarily obscures the pulsar.
Taking our fit results from above as basis (N\i{H,max}$\approx$2\mal
10\e{24}\,cm\e{-2}), the optical depth for Thomson scattering of such a
blob would be \ca 1.6, reducing the direct component to \ca 20\%. 
The scattered radiation would need to come from a relatively large region
(\ca 10\e{13}\,cm) to destroy coherence. The large fraction of
scattered radiation would also explain the relatively increased
Fe-line emission and soft excess.

After the quiescent state, when pulsations begin again, the emission in this
scenario would be a combination of heavily absorbed direct radiation -- the
accretion being fueled by some part of the blob -- and scattered radiation
from a wide region. This would also explain the reduced pulse fraction, due
mainly to a higher ``pedestal'' during the high state as compared to the
normal state at the beginning of the observation (see \Fig{fig:PP}).




\small
\parskip 0pt
\bibsep 0pt
\renewcommand{\refname}{\small\textbf{References}}
\bibliographystyle{jwaabib}
\bibliography{mnenomic,AA-abrv,vela}

\begin{thebibliography}{}

\bibitem[\protect\astroncite{Inoue et~al.}{1984}]{Inoue:84}
Inoue H., Ogawara Y., Ohashi T., et~al., 1984, PASJ 36, 709

\bibitem[\protect\astroncite{Kreykenbohm et~al.}{1999}]{Kreykenbohm:99}
Kreykenbohm I., Kretschmar P., Wilms J., et~al., 1999, A\&A 341, 141

\bibitem[\protect\astroncite{Mihara}{1995}]{Mihara:Dr}
Mihara T.,  1995,
\newblock {\em Ph.D. thesis\/}, University of Tokyo

\bibitem[\protect\astroncite{Nagase et~al.}{1986}]{Nagase:86}
Nagase F., Hayakawa S., Sato N.,  1986, PASJ 38, 547

\bibitem[\protect\astroncite{Raubenheimer}{1990}]{Raubenheimer:90}
Raubenheimer B.C.,  1990, A\&A 234, 172

\bibitem[\protect\astroncite{{van Kerk\-wijk} et~al.}{1995}]{vanKerkwijk:95}
{van Kerk\-wijk} M.H., {van Paradijs} J., Zuiderwijk E.J., et~al., 1995, A\&A
  303, 483

\end{thebibliography}

\end{document}